\documentclass[prd, twocolumn, amsmath, amsfonts, amssymb, showpacs,
nofootinbib, preprintnumbers]{revtex4}
\usepackage{epsfig,graphicx,bm,color}
\usepackage{url}
\begin{document}
\title{Investigating Gamma-Ray Lines from Dark Matter with Future Observatories}
\author{Lars Bergstr\"om}
\email{lbe@fysik.su.se}
\affiliation{{T}he Oskar Klein Centre for Cosmoparticle Physics, Department of Physics, Stockholm University, AlbaNova, SE-106 91 Stockholm, Sweden}
\author{Gianfranco Bertone}
\email{gf.bertone@gmail.com}
\affiliation{GRAPPA Institute, University of Amsterdam, Science Park 904, 1090 GL Amsterdam,
Netherlands}
\author{Jan Conrad}
\email{conrad@fysik.su.se}
\author{Christian Farnier}
\email{Christian.Farnier@fysik.su.se}
\affiliation{{T}he Oskar Klein Centre for Cosmoparticle Physics, Department of Physics, Stockholm University, AlbaNova, SE-106 91 Stockholm, Sweden}
\author{Christoph Weniger}
\email{weniger@mpp.mpg.de}
\affiliation{Max-Planck-Institut f\"ur Physik, F\"ohringer Ring 6, 80805
M\"unchen, Germany}

\preprint{MPP-2012-121}



\begin{abstract}
  We study the  prospects for studying line features in gamma-ray spectra with
  upcoming gamma-ray  experiments, such as HESS-II, the Cherenkov Telescope
  Array (CTA), and the GAMMA-400 satellite.  As an example we use the narrow
  feature at 130 GeV seen in public data from the Fermi-LAT satellite. We
  found that all three experiments should be able to confidently confirm or
  rule out the presence of this 130 GeV feature. If it is real, it should be
  confirmed with a confidence level higher than 5$\sigma$. Assuming it to be a
  spectral signature of dark matter origin, GAMMA-400, thanks to a projected
  energy resolution of about 1.5 \% at 100 GeV, should also be able to resolve
  both the  $\gamma\gamma$ line and a corresponding $Z\gamma$ or $H\gamma$
  feature, if the corresponding branching ratio is comparable to that into two
  photons. It will also allow to distinguish between a gamma-ray line and the
  similar feature resulting from internal bremsstrahlung photons. 
\end{abstract}

\maketitle
\newcommand{\ga}{\gamma}
\newcommand{\be}{\begin{equation}}
\newcommand{\ee}{\end{equation}}
\newcommand{\bea}{\begin{eqnarray}}
\newcommand{\eea}{\end{eqnarray}}
\newcommand{\ds}{{\sf DarkSUSY}}
\newcommand{\py}{{\sf PYTHIA}}
\newcommand{\code}[1]{{\tt #1}}

\hyphenation{}

\section{Introduction}
As the Large Hadron Collider (LHC) keeps accumulating data at high luminosity
(and soon at full energy), hopes are high that it will help elucidating the
nature of the particle making up around 23 \% of the energy density of the
universe, the {\it dark matter} particle \cite{bertone_book,reviews}.  So far,
no such new mass scale has been found, although the prediction from
supersymmetric (SUSY) models that the lightest Higgs boson should weigh less
than 130 GeV~\cite{lhc_higgs}, which seems to be confirmed by the detection
recently done at CERN's LHC, which gives a mass of the potential Higgs boson
of around 125 GeV.

As for dark matter candidates, only constraints on the parameter space of the
most popular extensions of the Standard Model, in particular Supersymmetry,
have been obtained \cite{SUSYfits}, but even if such candidates were to be
found, it will be hard to prove with LHC data only that they actually
constitute most of the dark matter in the Universe, as the required lifetime
of many times the age of the universe would seem impossible to verify in
accelerator experiments \cite{Bertone:2010at}.   
 
Fortunately, {\it direct} and {\it indirect} dark matter searches will provide
complementary information, possibly allowing a precise identification of dark
matter particles \cite{Bertone:2010rv,Bertone:2011pq}. Direct detection by
scattering of dark matter particles traversing the earth in ultra-pure
counting experiments has historically been the most advanced technique, but
indirect detection methods have recently received increased interest (see e.g.
Ref. \cite{bertone_book} for reviews). 

Indirect detection is based on the search for secondary photons, antimatter,
and neutrinos produced by the  annihilation or decay of dark matter particles.
For $\gamma$-rays coming from annihilations of dark matter particles in the
halo, Fermi-LAT has very successfully delivered bounds that have started to
probe into the parameter space of viable models, in line with pre-launch
expectations \cite{prelaunch}, in particular for dwarf spheroidal galaxies
\cite{fermidwarfs} and galaxy clusters \cite{fermiclusters}. 

Recently, a possible hint of a dark matter signal in the form of a narrow {\it
spectral line} or an {\it internal bremsstrahlung} (IB) feature, has been
found in analyses of public data from the Fermi-LAT satellite detector
\cite{bringmann, weniger} (see also \cite{tempel, su_fink}). The signal is too
weak to claim a discovery, but being of a type and at an energy where there is
no other known astrophysical explanation\footnote{The very fine-tuned pulsar
model from \cite{aharonian_pulsar} can be disregarded since the signal is
significantly extended.} it is important to further study this type of
signature in independent experiments.

We take in this paper, as an exercise, the existence of these recent
indications for a line or an IB bump seriously, and we discuss how this
effect, if real, would appear in a number of existing (Fermi-LAT
\cite{fermi-lat}, HESS-II \cite{hess-2}) and planned (CTA \cite{cta},
GAMMA-400 \cite{gamma-400}) $\gamma$-ray detectors. If the present indications
of a line structure in the Fermi-LAT public data would disappear, our results
should be useful for future indirect dark matter searches. 

In particular, we discuss the possibility of one or more associated lines
coming from the $Z\gamma$ and $H\gamma$ (with $H$ the Higgs boson)
annihilation channels in some models, and we investigate whether one can
separate a line signal from other spectral features like IB emission.  We will
show that with the upcoming detector HESS-II, and the proposed Cherenkov
Telescope Array (CTA) and the GAMMA-400 satellite these gamma-ray structures,
if real, indeed will be confirmed with much higher confidence. 

The paper is organized as follows: in Sec. \ref{sec:theory} we review the
spectral features   arising from dark matter annihilation and the tentative
detection of a feature in Fermi data.  In Section \ref{sec:prospects} we
discuss the prospects to detect such feature with future observatories
focusing on improvements in energy resolution and effective area.  In
\ref{sec:conclusions} we discuss our results and present our conclusions. 
  
\section{Indirect detection with Gamma-Rays}
\label{sec:theory}

\subsection{Basics}
Among the possible secondary particles, gamma-rays play a special role in
indirect dark matter searches, since they propagate without being absorbed or
deflected in the local Universe, in contrast, e.g., to the case of anti-matter
(see \cite{reviews}).

The expected dark matter generated gamma-ray flux from a cone with solid angle
$\Delta\Omega$ observed at earth can be written 
\be
 \label {eq_gamma}
  \frac{d\Phi_\gamma}{dE}=  \frac{\langle\sigma v\rangle}{2m_\chi^2}\sum_f
  \frac{dN^f_\gamma}{dE}\times  \frac{B}{4\pi}\int_{\Delta\Omega}\!
  d\Omega\int_{l.o.s.} d\lambda\, \rho_\text{dm}^2(\lambda),
\ee
where $\langle\sigma v\rangle$ is the total average annihilation rate,
$N_\gamma^f$ the number of photons produced in annihilation channel $f$,
$\rho_\text{dm}$ the smooth part of the dark matter density and $\lambda$ the
line-of-sight distance in the direction of observation. This expression has to
be convolved with the energy and angular resolution of the detector. 

The dark matter density along the line of sight, $\rho_\text{dm}(\lambda)$, is
at present unconstrained in the inner Galaxy, and it is therefore derived from
N-body simulations, and most commonly approximated with analytic fits like the
NFW \cite{nfw} or the Einasto profile \cite{einasto}, for which the integral
along the line of sight turns out to produce similar values.  In
the rest of the paper we make use of the Einasto profile,
\begin{align}
  \rho_{\text{dm}}(r) \propto \exp \left(-\frac{2}{\alpha_E}
  \frac{r^{\alpha_E} }{r_s^{\alpha_E}}\right)\;,
\end{align}
with $\alpha_E=0.17$ and $r_s=20\ \rm kpc$, normalized to
$\rho_\text{dm}(r_0)=0.4\ \rm GeV\ cm^{-3}$ at position of the Sun, $r_0=8.5\
\rm kpc$.

A boost factor $B$ is often included in the calculation, to parametrize our
ignorance on the enhancements of the annihilation signal that may be caused by
processes like: the presence of dark matter substructures; the so-called
Sommerfeld enhancement factor \cite{sommerfeld} (although the possibility of a
very large $B$ factor is constrained by Cosmic Microwave Background
observations \cite{cmb}) or strong deviations of the dark matter distribution
from the Einasto or NFW profiles. 

As it has been clear for quite some time, it is not possible to reliably
predict the size of the boost factor $B$, as it (besides the particle physics
uncertainties relating to the Sommerfeld enhancement) depends critically on
the density profile and spatial distribution of subhalos, a problem largely
beyond the capability of present-day simulations.  What we know is that
numerical simulations suggest that there should be fewer substructures in the
inner galaxy than in the outer parts, due to the process of merger and tidal
stripping in the Milky Way halo (see e.g.  \cite{springel,Pieri:2009je} and
references therein), and that baryons dominate the gravitational potential at
the centre of the Galaxy, where galactic baryons such as the bar and the
central supermassive black hole strongly modify the distribution of dark
matter \cite{GC}. 

The dependence of the annihilation rate on the square of the dark matter
density may cause "hot spots" near density concentrations such as dwarf
galaxies, although recently it has appeared that galaxy clusters may be even
more interesting targets for detection claims \cite{clusters}. However, the
strongest signal should exist near the Galactic centre, as predicted by all
halo modeling, including N-body simulations, and we will concentrate on that
target throughout this work.

\subsection{Identity of a $\gamma$-ray line signal}
Especially important for the clear demonstration of a dark matter signal is
the  possibility of ``smoking gun'' signatures like gamma-ray lines
\cite{lines,zgamma} or similar structures caused by internal bremsstrahlung
(IB) \cite{fsr}. Presently no known astrophysical background processes that
can give such conspicuous signals have been observed.

\begin{figure}[t]
  \begin{center}
    \includegraphics[width=\columnwidth] {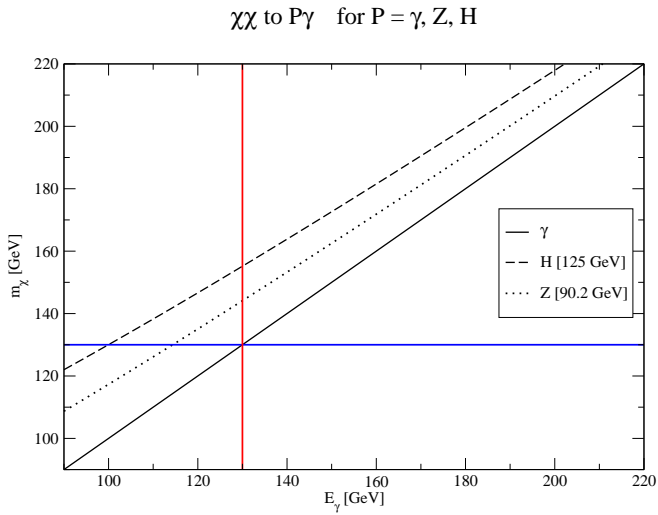}
  \end{center}   
  \caption{Illustration of Eq.~(\ref{eq:altlines}), which relates the mass
  $m_\chi$ of the annihilating particles with the energy of the $\gamma$-ray
  line,for the case $\chi\chi\to P\gamma$, with $P=\gamma, H, Z^0$. Here $m_Z$
  is 90.2 GeV, and as an illustration the Higgs mass has been set to 125 GeV,
  indicated by present LHC data \protect\cite{lhc_higgs}.} 
  \label{fig:lines}
\end{figure}

Depending on the structure of the theory of the dark matter particle $\chi$,
one can imagine at present at least three different origins of a sizeable
narrow line. The most obvious (and in many cases the strongest) will come from
the $\chi\chi\to \gamma\gamma$ process, which will give a line at energy
$E_\gamma=m_\chi$, with a very narrow intrinsic width that will be dominated
by Doppler motion of the annihilating pair, $\epsilon_\gamma\equiv\delta
E_\gamma/E_\gamma\simeq 10^{-3}$ for typical Galactic velocities. For
annihilation into another neutral particle plus a photon, like the electroweak
gauge boson $Z^0$ or the Higgs boson $H$, the line width will be dominated by
the intrinsic width of the decay particle, which for the $Z^0$ case will give
$\epsilon_\gamma\simeq 10^{-2}$, and smaller for a standard model Higgs boson,
if it coincides with the detection claimed at LHC.  The relation between the
photon energy and the mass of the particle $P$ in the final state is
\begin{equation}
  E_\gamma=m_\chi\left(1-\frac{m_{P}^2}{4m_\chi^2}\right)
\end{equation}
or, solving instead for $m_\chi$
\begin{equation}
  m_\chi=\frac{1}{2}\left(E_\gamma+\sqrt{m_P^2+E_\gamma^2}\right)\;.
  \label{eq:altlines}
\end{equation}
This relation is shown in Fig.~\ref{fig:lines}. Assuming that the observed
gamma-ray line at 130 GeV is due to the $\chi\chi\to \gamma\gamma$ process one
finds, following the horizontal line, predictions for the location of
$H\gamma$ and $Z^0\gamma$ lines at 100 GeV and 114 GeV, respectively.
Alternatively, following the vertical line, one sees that if the observed 130
GeV line is a result of the $\chi\chi\to H\gamma$ or $Z^0\gamma$ process, the
$\chi$ mass is 155 or 142 GeV, respectively.

\vskip .5cm

\begin{table}[hbt!]
  \begin{center}
    \begin{tabular}{|p{1.5cm}|p{1.5cm}|p{1.5cm}|p{1.5cm}|}
      \hline  & $\gamma\gamma$ & $H\gamma$ & $Z\gamma$\\ 
      \hline $\gamma\gamma$ & 130 & 100 & 114\\
      \hline $H\gamma$ & 155 & 130 & 142\\
      \hline $Z\gamma$ & 144 & 117 & 130\\
      \hline
    \end{tabular}
  \end{center}
  \caption{Predicted $\gamma$-ray energies, in GeV, if the 130 GeV line
  originates from the process indicated by the row, for the process given by
  the respective column.}
  \label{tab:tab1}
\end{table}
\vskip .5cm     
 
The predicted energies of all three possible lines, the ones coming from
$\gamma\gamma$, $H\gamma$ and $Z^0 \gamma$ (with $m_H$ set to 125 GeV), for
all permutations are shown in Table~\ref{tab:tab1}. It will depend on the
model if all three lines are allowed. In particular, as a radiative $0\to 0$
transition is forbidden due to gauge invariance and angular momentum
conservation, the annihilation to $H\gamma$ is not allowed from the dominant
$s$ wave in the Galaxy if $\chi$ is a Majorana fermion or a spin-0 particle
\cite{lbereview}. For definiteness, we will in the following assume that the
tentative 130 GeV structure is due to the $\gamma\gamma$ line, but we will
also compare with the expectations for the IB effect, to which we now turn.  

\subsection{The internal bremsstrahlung effect}
The $\gamma\gamma$ process normally appears in a closed loop containing the
various charged particles to which the dark matter particles couple. This
means that it is generally suppressed by powers of the electromagnetic
coupling constant, and the cross section will contain an explicit factor
$\alpha^2_{em}$.  An interesting effect appears, however, for Majorana
fermions already at order $\alpha_{em}$. It was early realized that there
could be  important spectral features \cite{lbe89}, and recently it has been
shown that internal bremsstrahlung (IB) from charged particles in the
$t$-channel in the annihilations could yield a detectable, quite sharp "bump"
near the highest energy, i.e., at the rest mass of one of the annihilating
particles moving slowly ($v/c\sim 10^{-3}$) in the Galactic halo
\cite{bbe,bringmann,fsr}.  In \cite{birkedal}, it was furthermore pointed out
that  final state radiation (FSR) often can be estimated by simple, universal
formulas and often gives rise to a very prominent step in the spectrum at
photon energies of $E_\gamma=m_\chi$.  The IB and FSR processes was thoroughly
treated in \cite{bbe} (see also \cite{bringmann,fsr}), and here we summarize
the main results. 

In Ref.~\cite{lbe89} it was shown that the  radiative process $\chi^0\chi^0\to
f\bar f\gamma$  may circumvent the chiral suppression, i.e., the annihilation
rate being proportional to $m_f^2$. This is normally what one would get for
annihilation into a fermion pair from an $s$-wave initial state
\cite{goldberg}, as is the case in lowest order for non-relativistic dark
matter Majorana particles in the Galactic halo.  A fermion final state with an
additional photon, $f\bar f\gamma$, is thus surprisingly not subject to  a
helicity suppression. The full analytical expressions are lengthy, but
simplify in the limit of $m_f\rightarrow0$. Then one finds in the
supersymmetric case \cite{bbe} for the radiative differential rate, normalized
to the $f\bar f$ rate 
$$
\frac{dN_f^{\gamma,\mathrm{IB}}}{dx} =\hspace{5.5cm}
$$
\be
\Delta\times\left[\frac{4x}{\mu(\mu-2x)}
 -\frac{2x}{(\mu-x)^2}
   -\frac{\mu(\mu-2x)}{(\mu-x)^3}\log\frac{\mu}{\mu-2x}\right]\,, \label{eq:ib}
\ee
with 
$$ 
\Delta = (1-x) \alpha_\mathrm{em}Q^2_f\frac{\left|\tilde
g_R\right|^4+\left|\tilde g_L\right|^4}{64\pi^2} \left[m_\chi^2 \langle\sigma
v\rangle_{\chi\chi\rightarrow f\bar f}\right]^{-1}\;,
$$
where $Q_f$ is the electric charge of the fermion, $\mu\equiv m_{{\tilde
f}_{L,R}}^2/m_\chi^2+1$ and $\tilde g_RP_L$ ($\tilde g_LP_R$) is the coupling
between the dark matter particle, the fermion and right-handed (left-handed)
$t$-channel exchange particle $\tilde f$ (which would be a sfermion in the
supersymmetric case). This agrees with the result found in \cite{lbe89} for
the case of pure photino annihilation (see  \cite{baltz_bergstrom} for a
right-handed neutrino dark matter candidate where this interesting internal
bremsstrahlung effect also appears). Note the large enhancement factor
$m_\chi^2/m_f^2$ due to the lifted helicity suppression (from ${\langle\sigma
v\rangle}_{\chi\chi\rightarrow f\bar f}\propto m_f^2m_\chi^{-4}$), and another
large enhancement that appears at high photon energies for sfermions
degenerate with the neutralino.

For the supersymmetric case, internal bremsstrahlung from the various possible
final states of neutralino annihilations is included in \ds\ \cite{darksusy}.
The total $\gamma$-ray spectrum  is given by \be\label{eq:tot}
\frac{dN^{\gamma,\mathrm{tot}}}{dx}=\sum_f
B_f\left(\frac{dN_f^{\gamma,\mathrm{sec}}}{dx}+\frac{dN_f^{\gamma,\mathrm{IB}}}{dx}
+ \frac{dN_f^{\gamma,\mathrm{line}}}{dx}\right)\,,
\ee
where $B_f$ denotes the branching ratio into the annihilation channel $f$. The
last term in the above equation gives the contribution from the direct
annihilation into photons, $\gamma\gamma$ or $Z\gamma$, which as mentioned
before result in a sharp line feature \cite{lines,zgamma}. The first term is
the contribution from secondary photons from the fragmentation of the fermion
pair.  This ``standard'' part of the total $\gamma$-ray yield from dark matter
annihilations shows a feature-less spectrum with a rather soft cutoff at
$E_\gamma=m_\chi$.

\begin{figure}[t] 
  \begin{center} 
    \includegraphics[width=8cm]{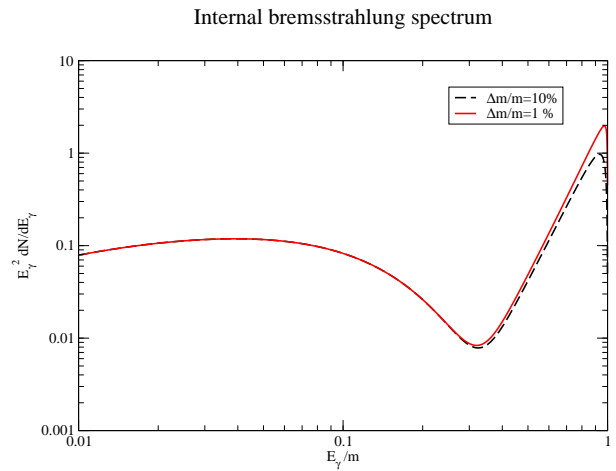}
  \end{center} 
  \caption{The distribution of $\gamma$-rays from the internal bremsstrahlung
  process $\chi\chi\to f\bar f\gamma$ (see \protect\cite{bringmann}), for two
  different values of the relative mass difference between the $t$-exchange
  particle and the dark matter Majorana fermion $\chi$. (Here a continuous
  spectrum of the $b\bar b$ type is also included.) As can be seen, the
  internal bremsstrahlung process gives a very hard spectrum which peaks near
  the kinematical end point, and thus is a ``smoking gun'' signature for dark
  matter annihilation, similar to the $\gamma$-ray line signal.}
  \label{fig:ib}
\end{figure}

In Fig.~\ref{fig:ib} an example of the energy distribution of photons given by
the first two terms in (\ref{eq:tot}) is shown. To this may be added one or
more line signals, the strength of which is very model dependent, however. 

\subsection{Brief review of present indications of {\boldmath$\gamma$}-ray line
signal} 

Past searches for gamma-ray lines in the Fermi-LAT or EGRET data found no
indications for line signatures below 200 GeV, and presented upper limits on
the annihilation cross-section (or decay rate, in some models) of dark matter
particles into monochromatic photons~\cite{egret, fermilines, vertongen,
ferminew, fermiextra}. 

However, recent analyses of public Fermi-LAT data \cite{fermi-lat} have
identified a feature at 130 GeV in a line-search from the halo in the vicinity
of the galactic centre, that has tentatively been interpreted in terms of dark
matter  \cite{bringmann,weniger}. Beside taking into account the full set of
available data, the major improvement with respect to previous studies was an
adaptive selection of search regions with optimized signal-to-noise level for
different profiles of the Galactic dark matter halo. 

The revealed signature is too sharp to allow a straightforward astrophysical
explanation, and strongly suggestive towards an interpretation in terms of a
gamma-ray line at 130 GeV~\cite{weniger}, or alternatively as the virtual
internal bremsstrahlung from dark matter particles with a mass of around 150
GeV~\cite{bringmann}. The signature appears close to the Galactic centre only.

This discovery has already gained considerable attention (see, e.g.,
\cite{linepapers}).  Ref.~\cite{profumo_linden} suggested the Fermi Lobes
(bubbles/haze)~\cite{bubbles} as cause for the observed signature; but this
possibility was however quickly disputed by Ref.~\cite{tempel} and
Ref.~\cite{su_fink}, who emphasized the locality of the signature close to the
Galactic centre and relaxed the power-law approximation to the background
fluxes. 

In particular, in the analysis in \cite{su_fink} a signal of more than
$5\sigma$ was found, but caveats of the interpretation are also given: there
is a modest evidence of an $1.5\,^{\circ}$ offset of the signal region with
respect to the Galactic centre (this may however be explained by the interplay
between baryons and dark matter \cite{kuhlen}), and there are weak indications
of a possible instrumental effect giving spurious lines of which one also
happens to be at 130 GeV. In either case, whether these new results are a
consequence of a genuine dark matter signal, or point to an unexpected
instrumental effect of the Fermi-LAT detector, it is imperative that this
signal region is probed by independent detectors.

If interpreted in terms of dark matter annihilation, the signature is
consistent with an Einasto or NFW profile~\cite{weniger, su_fink}. In this
case, the annihilation cross-section for $\chi\chi\to\gamma\gamma$ is best
given by $\langle\sigma v\rangle \simeq1.3\cdot 10^{-27}\rm cm^3s^{-1}$, with
a dark matter mass of $m_\chi\simeq 129.8\pm
2.4\,^{+7}_{-13}$~GeV~\cite{weniger}, or $127.0\pm 2.0$~GeV~\cite{su_fink}.

\section{Prospects for identifying spectral features with future observatories}
\label{sec:prospects}

\subsection{Statistical treatment} 
In this work, we study the prospects for detecting and identifying sharp
gamma-ray signatures (possibly arising from dark matter annihilation) with
GAMMA-400, HESS-II and CTA, by adopting the standard methods that were used in
previous line searches~\cite{egret, fermilines, vertongen, ferminew, weniger}
and apply them to different sets of appropriately generated mock
data.\footnote{A complementary approach that could be adopted in the analysis
of GAMMA-400 data would be a spatial analysis as in Ref.~\cite{su_fink}.} For
each instrument under consideration, the basic strategy is: (1) Determine a
region of the sky with large expected signal-to-noise ratio; this is typically
the Galactic centre, maybe plus regions above/below the Galactic disc. (2)
Perform a spectral analysis of the (here simulated) gamma-ray fluxes in this
region using the maximum likelihood technique~\cite{wilksTheorem}.

Throughout, best-fit model parameters are inferred by maximizing the
likelihood function $\mathcal{L}=\Pi_i P(c_i|\mu_i)$, where $P(c|\mu)$ is the
Poisson probability to observe $c$ events when $\mu$ are expected, $c_i$
denotes the number of events that fall into energy bin $i$, and $\mu_i$ is the
number of expected events in bin $i$ and a function of the model parameters
and the instrument response function. All fits are only performed within small
energy windows centred around the line position(s). The small energy windows
allow to approximate astrophysical background fluxes locally by a single
power-law (we leave normalization and spectral slope free in the fits), and it
eliminates the need to include secondary signal photons into the fits (because
they would peak far outside the window).  The energy window ranges per
definition from $E_\gamma/\sqrt{\epsilon}$ to $E_\gamma\sqrt{\epsilon}$;
$E_\gamma$ is the line energy, the energy window size $\epsilon$ is typically
a factor of a few larger than the energy resolution of the respective
instrument and will be defined below. Note that in all fits we keep the
position of the line-like features fixed.  Our treatment of the instrument
response functions that we need in order to obtain $\mu_i$ are described
below; fits are performed using the \texttt{ISIS} package~\cite{isis}.

Projected upper limits at the $95\%$CL are derived using the profile
likelihood method~\cite{pro_like}: We generate a large number of mock data
sets without a signal. We then fit the mock data with a power-law with free
slope and normalization plus a line signal with free non-negative
normalisation and increase the line flux from its best-fit value until
$-2\ln\mathcal{L}$ changed by 2.71 (while profiling over the background
parameters). For different line energies, we quote the log-scale mean of the
limits obtained this way as the expected limit.

\begin{figure*}[!htb]
  \begin{center}
    \includegraphics[width=0.8\textwidth] {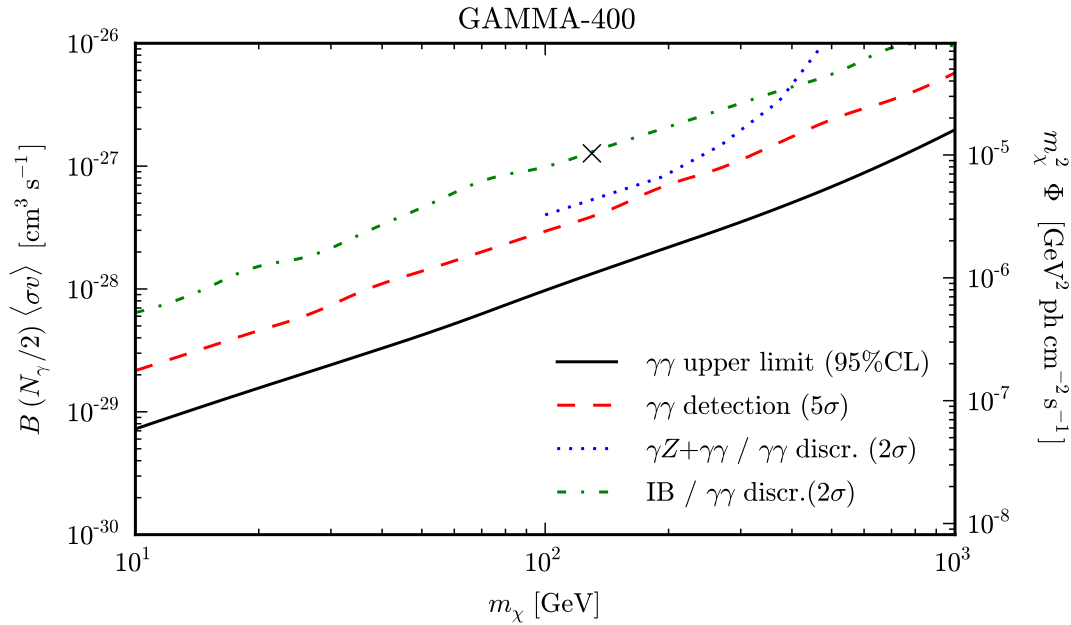}
  \end{center}   
  \caption{Estimated flux and velocity weighted cross-section sensitivities
  obtained for GAMMA-400 as a function of the DM particle mass. \emph{Black
  solid:} Expected upper limits for $\gamma\gamma$ final states (with
  $N_\gamma=2$); \emph{red dashed:} detection of $\gamma\gamma$ at $5\sigma$;
  \emph{green dot-dashed}: discrimination between IB and a monochromatic line
  ($N_\gamma=1$); \emph{blue dotted:} discrimination between $\gamma
  Z+\gamma\gamma$ and $\gamma\gamma$ (assuming
  BR($\chi\chi\to\gamma\gamma$)=BR($\chi\chi\to\gamma Z$)=0.5;
  $N_\gamma=1.5$). The cross indicates the 130 GeV signature from
  Ref.~\cite{weniger}.}
  \label{fig:lines_gamma400}
\end{figure*}

The local significance for a line signature can be derived using the
maximum-likelihood-ratio test; it is given by the square-root of the TS-value
$TS\equiv-2\ln(\mathcal{L}_{\rm null}/\mathcal{L}_{\rm alt})$, where
$\mathcal{L}_{\rm null}$ and $\mathcal{L}_{\rm alt}$ are the likelihoods of
background-only (a single power-law) and background-plus-line fits,
respectively. For the different experiments, we determine the $5\sigma$
detection threshold for monochromatic photons by adjusting the signal flux
such that the TS-value averaged over a large number of mock data sets
is~23.7~\cite{Chernoff}.

In case a narrow line is identified, it is imperative to devise a strategy to
discriminate between IB and monochromatic photons, and to test for the
presence of additional closeby lines. We are interested in the flux at which
such a discrimination becomes possible at the $95\%$CL. To this end, we
simulate data with IB features or with two lines (corresponding to
$\gamma\gamma$ and $\gamma P$ final states, with $P=\rm H, Z$), and perform a
maximal likelihood ratio test to compare the likelihood of IB/two-line fits
with the one-line scenario.  We search for the flux for which the average of
the log-likelihood ratio is $\langle-2\ln(\mathcal{L}_{\rm
one-line}/\mathcal{L}_{\rm two-lines/VIB})\rangle=4$.  In case of the two-line
model, we leave the relative normalization of the two lines, as well as their
position, fixed. The IB signal from dark matter with mass $m_\text{dm}$ is
compared with a monochromatic line at $E_\gamma \approx 0.90\ m_\text{dm}$,
since the IB feature peaks slightly below the kinematical threshold; in these
fits the dark matter mass is left as a free parameter.  Since the models are
not nested, the interpretation of the log-likelihood ratio is in principle not
straightforward.  However, the obtained significance levels are conservative
with respect to what is obtained in a comprehensive likelihood
test~\cite{james}; for a study of the comprehensive likelihood method in the
context of dark matter searches see \cite{Conrad2012}.

\subsection{Effects of energy resolution -- The case of GAMMA-400}

The search of a weak gamma-ray line-like signal on top of a continuous
background spectrum may be hindered by statistical fluctuations of the
background. This is particularly true if the signal is additionally spread
over a large energy range due to the limited energy resolution of the
instrument. For pair conversion gamma-ray space telescopes, one of the main
instrumental characteristics influencing the resolution at high energies is
the calorimeter dimensions, measured in radiation lengths ($X_0$). This is
because the energy reconstruction at high energies is crucially depending on
how well the electromagnetic shower can be mapped. Fermi-LAT with its 8
radiation lengths reaches an energy resolution of around $\sim$ 10 \% at
energies of about 100 GeV. 

The design of the GAMMA-400 experiment, a future Russian-Italian pair
conversion telescope whose preliminary launch date is announced to 2018
\cite{gamma-400}, should be similar to that of the Fermi-LAT. It will consist
of a tracker unit, a calorimeter and anti-coincidence systems. Design details,
such as number of tracker layers, presence of converter material in the
tracker  and design of anti-coincidence systems (which will be very demanding
due to a very excentric orbit) are not fixed yet, but it is very probable that
GAMMA-400  will be equipped with a 22 $X_0$ calorimeter with claimed energy
resolution of about 1.5 \% at 100 GeV. 

The effective area of GAMMA-400 is however foreseen to be smaller (around
$6400\ \rm cm^{2}$) above a few tens of GeV and constant. These two design
features immediately allow a rough estimate of the expected improvement: for
upper limits $\sqrt{2/10} \sim 0.4$, for signal significance $10/2 \sim 5$.
Interestingly, a larger calorimeter also will help with background rejection
(which is largely based on shower shape) and extend the usable energy range to
larger energies.  As mentioned, the design phase is not concluded and the
design still allows considerable freedom. Here, we assume the energy
resolution of GAMMA-400 to scale as the one of the Fermi-LAT (such that it
ranges from $3\%$ at 10 GeV to $1\%$ at 1 TeV).

\begin{figure*}[!htb]
  \begin{center}
    \includegraphics[width=0.7\textwidth] {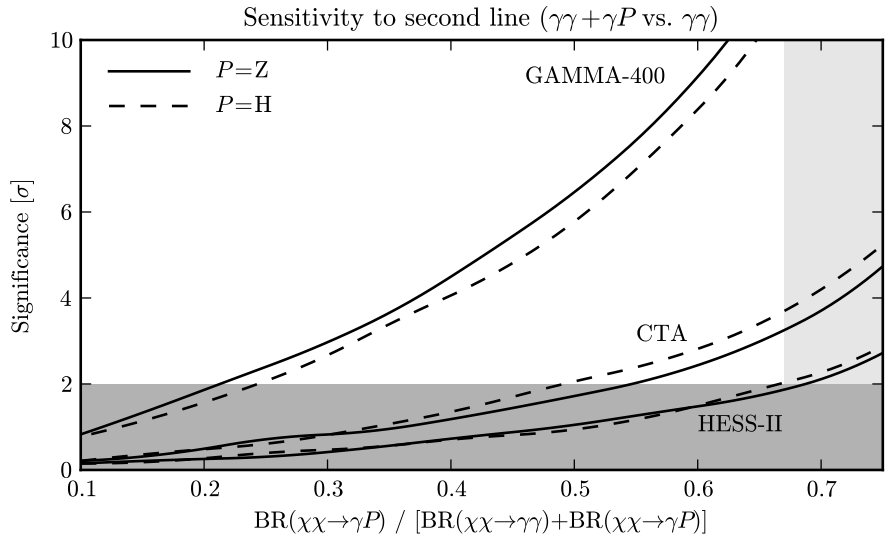}
  \end{center}   
  \caption{Assuming that the 130 GeV feature from Ref.~\cite{weniger} is due
  to annihilation into $\gamma\gamma$, we show as function of the branching
  ratios the significance at which a second line (either $\gamma \rm Z$ or
  $\gamma \rm H$) could be detected by different experiments. In the right
  gray area the secondary line would be as strong as the primary one; in the
  lower gray area the significance would fall below $2\sigma$.}
  \label{fig:brScan}
\end{figure*} 

With the hypothesis that the final design of GAMMA-400 is similar to the one
described above, we derived the sensitivity of GAMMA-400 for spectral
signatures from dark matter annihilation after 5 years of full sky survey mode
operation. As target region, we select a circular region around the Galactic
centre with $20^\circ$ radius excluding the Galactic disc part ($|\ell| >
5^\circ\ \&\ |b|<5^\circ$). To model the background that will be observed by
GAMMA-400, we make use of the current observations of the gamma-ray emission
observed by the Fermi-LAT as reported in~\cite{weniger}; the events observed
above few GeV in our region of interest can be described by a power-law
function
$\mathrm{dN}/\mathrm{dE}_{\mathrm{bckg}} = 6 \cdot 10^{-11} \times \left (
\frac{E}{1GeV} \right )^{-2.5}
\mathrm{ph}\,\mathrm{GeV}^{-1}\mathrm{cm}^{-2}\mathrm{s}^{-1}$. The exposure
time of the source region is $3.2 \times 10^{7} \mathrm{s}$, assuming a
field-of-view of 2.4~sr. In our spectral fits, we will adopt the energy
windows from Ref.~\cite{weniger}, ranging from $\epsilon=1.5$ at 10 GeV to
$\epsilon=3.1$ at 1 TeV. We checked that our results do not critically depend
on this choice.\smallskip

In Fig.~\ref{fig:lines_gamma400}, our results for the projected $95\%$CL upper
limits from GAMMA-400 are shown by the black solid line. Even if the effective
area is smaller than the one of Fermi-LAT, the better energy resolution of the
instrument allows to better distinguish a deviation due to a line signal from
the power law background fluctuations. This would improve by a factor about
two to three the upper limits that are obtained by Fermi-LAT in the same
observational time.

We also computed the strength of a line-like signal in order to obtained a
5$\sigma$ level detection for GAMMA-400 after 5 years of survey mode. In this
case, the $\gamma\gamma$ annihilation cross-sections range between $ 10^{-29}
\,\mathrm{and}\, 5 \cdot 10^{-27} \mathrm{cm}^{3}\mathrm{s}^{-1}$, depending
on the mass of the DM particle (see red dotted line on
Fig.~\ref{fig:lines_gamma400}). The confirmation of the tentative 130~GeV line
at $5\sigma$ would require $\sim20$ months in survey mode, and $\sim10$ months
using pointed observation.

For such good energy resolution performances, the instrument should also be
able to probe efficiently the existence of a secondary line. As an
illustrative example, we consider here the case that dark matter annihilates
into $\gamma\gamma$ and $\gamma Z$ final states with the same rate. For such
scenarios, GAMMA-400 will be able to distinguish a two-line model from a
single line one at $95\%$ CL in the region indicated by the blue dotted line
in Fig.~\ref{fig:lines_gamma400}.  At energies above few hundred GeV, both
lines start to overlap, and a distinction becomes impossible.

\begin{figure*}[!t]
  \begin{center}
    \includegraphics[width=0.8\textwidth] {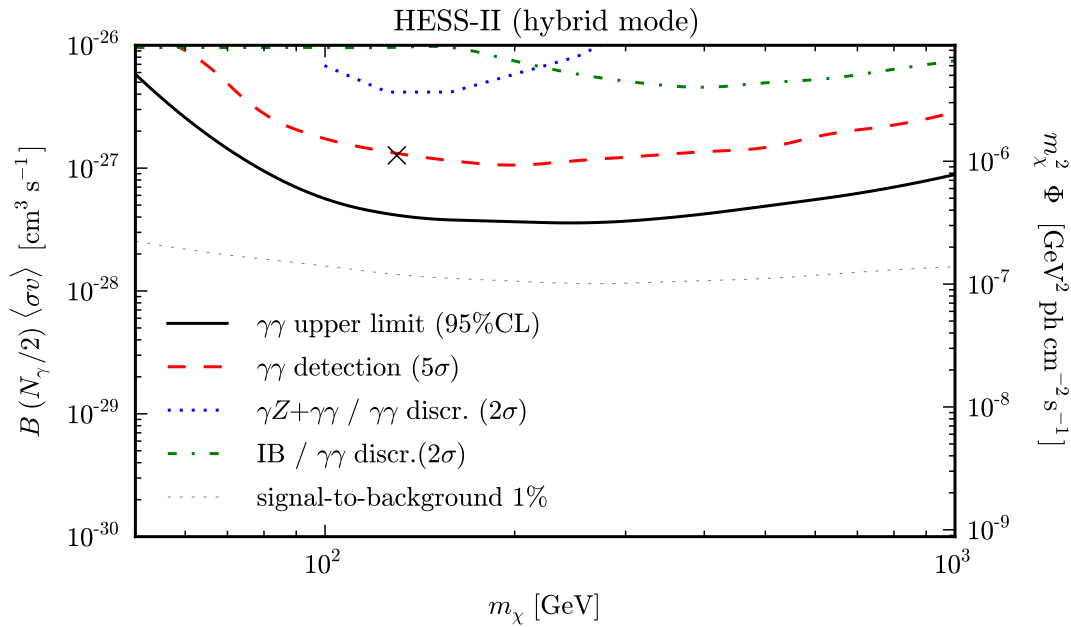}
  \end{center}   
  \caption{Same as Fig.~\ref{fig:lines_gamma400}, but for 50~h of
  Galactic center observation with HESS-II. The gray dotted line shows
  cross-sections at which the signal-to-background ratio at the line peak
  would drop below $1\%$.}
  \label{fig:lines_HESSII}
\end{figure*} 

Assuming that the 130 GeV feature from Ref.~\cite{weniger} is produced by
annihilation into $\gamma\gamma$ final states, we calculated the significance
at which GAMMA-400 will be able to identify secondary lines from either
$\gamma \rm Z$ or $\gamma \rm H$ (see Tab.~\ref{tab:tab1}) for different
branching ratios. In Fig.~\ref{fig:brScan}, we show the expected significance
as function of the branching ratio into the secondary line. As one can see
from this plot, GAMMA-400 would be an important probe for additional lines,
down to BR($\chi\chi\to\gamma P$) / BR($\chi\chi\to\gamma\gamma$)$\sim0.2$.

Finally, we investigate for which signal fluxes a signal arising from the
annihilation into $\gamma\gamma$ can be distinguished from IB photons. We fix
here the mass splitting (see above) to $\mu=2.1$ for definiteness. In
Fig.~\ref{fig:lines_gamma400}, we show by the green dash-dotted line the
annihilation cross-sections for which GAMMA-400 would be able to distinguish
at a $95\%$CL between monochromatic photons and IB photons. In particular, a
study of the nature of the 130 GeV excess seems realistic.

\subsection{Effects of increased area -- The case of CTA and HESS-II}

Space-based experiments are generally strongly size limited. The Fermi-LAT is
currently the largest pair-conversion telescope ever launched, and its
effective area is below $1\ \rm m^2$. This particularly reduces the
possibility of gamma-ray astronomy in the TeV domain, since the number of
observed events would be less than one per year even for the strongest steady
source. TeV astronomy is therefore performed with ground-based experiments,
where the secondary particles produced by gamma-rays in the atmosphere are
imaged and thereby energy and direction of the incident gamma-ray are
reconstructed. The effective area is therefore determined by the typical area
covered on the ground, resulting in improvements (w.r.t to pair conversion
telescopes) by about 4 orders of magnitude at 1~TeV.

Imaging the showers is performed either by Cherenkov detectors on ground
(water Cherenkov detectors such as MILAGRO) or by Imaging Air Cherenkov
Telescopes. Both techniques reach large effective areas on expense of
increased energy threshold, which for the former is around a TeV or the later
between 50 and 100 GeV. For the signal in question here therefore IACTs are
more relevant. Currently, they typically consist of 2 to 5 large size
telescopes (mirror diameter $\sim $ 10-20~m), equipped with fast electronic
camera recording the very brief and dim bunch of Cherenkov photons produced
during the secondary particles shower development in the atmosphere. Several
experiments are currently in operation: VERITAS in the United States and MAGIC
at La Palma cover the northern hemisphere, whereas H.E.S.S., located in
Namibia, is the best suitable experiment to observe the Galactic centre with a
low energy threshold. The performance of the 4 telescopes configuration,
operational since 2004, is too poor at energies relevant for the potential
signal. However, the addition of a much larger fifth ($600\ \rm m^2$)
telescope, being currently installed (denoted HESS-II) will allow to record
showers initiated by incident particles of lower energy. This is therefore the
configuration we focus on.  Preliminary estimates of the effective area and
energy resolution of HESS-II for an hybrid mode\footnote{The hybrid-mode
consists of the observation of the same events by the largest telescope and at
least one of the 13~m diameter ones.} operation have been reported in
\cite{hess-2}.  This operating mode will allow to observe events above 50~GeV
with an energy resolution of about 20\%.

\begin{figure*}[!t]
  \begin{center}
    \includegraphics[width=0.8\textwidth] {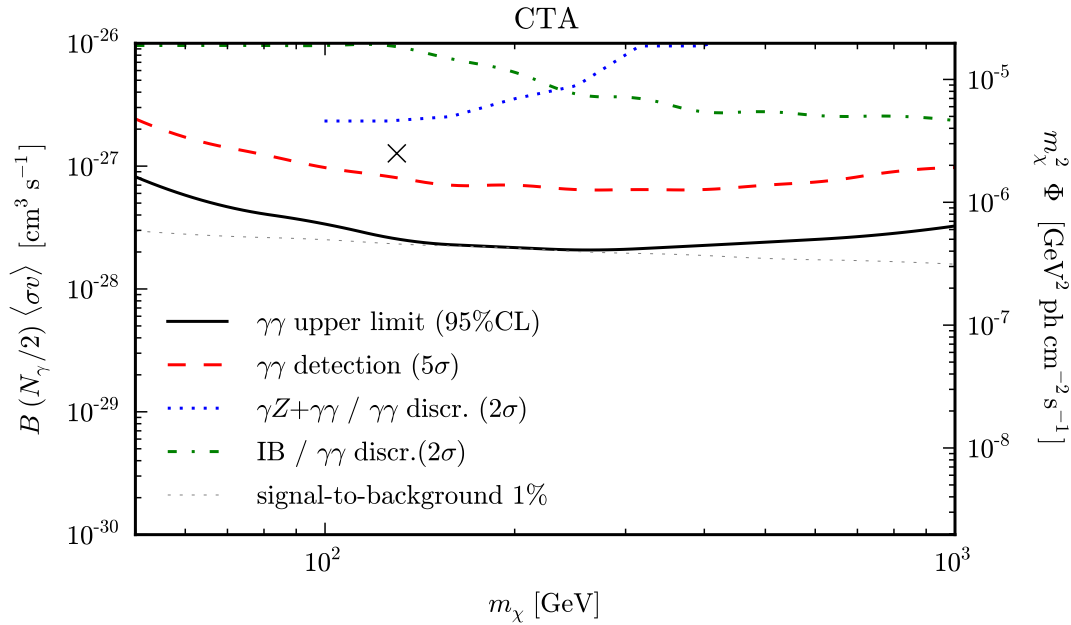}
  \end{center}   
  \caption{Same as Fig.~\ref{fig:lines_gamma400}, but for 5~h of
  Galactic center observations with CTA.}
  \label{fig:lines_CTA}
\end{figure*}

For IACT experiments, several sources of background have to be taken into
account.  First of all, charged cosmic rays, hitting the Earth with a much
greater rate than the most powerful gamma-ray sources, can produce Cherenkov
showers mimicking the signal induced by VHE gamma-rays. This is especially
true for electron induced showers which displayed very similar properties.
Hadronic showers, mainly due to protons, often carry secondary particles with
larger transverse momentum with respect to the shower direction, which
translate into more scattered images reconstructed in the focal plane of the
antennas.  Hence only a fraction ($\mathcal{O} (10^{-2})$) of these events
remain.

\medskip

For the signal region, in case of HESS-II, we adopt a circular target region
of $1^\circ$ radius around the Galactic centre. As energy window, we adopt two
times the window size in Ref.~\cite{1106.1874}. For observations toward the
Galactic centre, two additional gamma-ray signals have to be taken into
account: the detected point-like HESS source HESS~J1745-290
\cite{hess1745-290} and the diffuse emission around the Galactic centre also
measured by HESS \cite{hess_diffuse}.  For these four components, we used a
similar description as the one described in Appendix A of \cite{1106.1874}. 

As for GAMMA-400, we investigated the capabilities of HESS-II to detect a
gamma-ray line-like DM signal. We assumed an observation of 50~h towards the
Galactic centre, for intermediate zenith angle (20$^\circ$). The 95\%
C.L.~achieved with HESS-II operating in hybrid-mode, should be as low as few
$10^{-28} \mathrm{cm}^{3}\mathrm{s}^{-1}$ for DM masses between 80 to
1000~GeV, see figure~\ref{fig:lines_HESSII}. Since the energy resolution of
HESS-II is almost a factor 10 worse than what should be achieved with
GAMMA-400, we clearly see the benefits of a large number of events. However,
such energy resolution drastically reduce the possibility to distinguish a
two-lines scenario.  Nevertheless, like GAMMA-400, HESS-II will be able to
confirm the excess reported in \cite{weniger}.  \medskip

In the near future, Cherenkov Telescope Array (CTA) will be the next
generation of imaging atmospheric Cherenkov telescopes. It aims at increasing
the sensitivity of the current experiments up to a factor $\sim$ 10 in the TeV
domain, and also to lower the energy threshold to few tens of GeV. The current
design of the full observatory holds $\geq$ 60 telescopes, for which nearly
2/3 will be installed in the southern hemisphere for a better view of the
Galactic centre, see \cite{cta} for a detailed description of the design and
estimates of the performances of the instrument. We use the energy dependence
of the energy resolution of \cite{cta}, and we also assume the energy
dependence of the effective area from \cite{cta-icrc}. As signal region, we
adopt a circular target region of $2^\circ$ around the Galactic center.

The signal region definition and background description we adopted for CTA are
similar to the ones discussed above for HESS-II. With the presence of several
large size telescope in the centre of the array, CTA will be more sensitive
than HESS-II. In order to remain in the statistically limited regime, we
reduced the observing time for CTA down to 5 hours. At very low energies,
since we impose that the showers are observed at least by two telescopes, CTA
in its current design will not have a much larger effective area. Moreover the
intrinsic fluctuations in the low energy showers limit the energy resolution
that will be achieved even with several telescopes, so that in this energy
range, we do not expect stronger limits from CTA. Above 80~GeV however, the
higher telescopes multiplicity will increase the performance (larger effective
area and better energy resolution) so that CTA sensitivity will be almost
constant up to 1~TeV.  From figure~\ref{fig:lines_CTA}, we also see that CTA
will be very important to probe line-like signals from DM annihilations above
100~GeV, since after 5~h of observations of the Galactic centre, CTA will be
more sensitive than five years of GAMMA-400 observations.

\section{Discussion and Conclusions}
\label{sec:conclusions}

The detection of a sharp feature at an energy of 130 GeV in Fermi-LAT data has
sparked the interest of the astroparticle community, since the presence of
gamma-ray lines has long been considered a smoking-gun signature of new
physics, possibly pointing to the annihilation of dark matter particles. Of
course, future Fermi-LAT data will be very important: If the Fermi-LAT
collaboration can exclude instrumental effects as the cause of the structure,
it may well, in case upcoming data strengthens the feature, confidently
establish discovery of the effect. In any case, future gamma-ray observatories
would provide necessary independent confirmation and are expected to clarify
the experimental situation, in view of their increased effective area or
better angular resolution. In particular we focused here on three upcoming
experiments: HESS-II, CTA and GAMMA-400.

We summarize here the main results: 

\begin{itemize}
  \item We have calculated the sensitivity to gamma-ray lines for the three
    experiments, and we have shown that all of them will be able to confirm or
    rule out the presence of the 130 GeV line. In all cases, in fact, the
    feature found in Fermi-LAT data would be detectable with a significance
    higher than 5$\sigma$.
  \item We have assessed, for each experiment, the prospects for identifying
    the presence of additional lines, which would allow a better
    reconstruction of the particle properties of the annihilating dark matter
    particle. We found that only GAMMA-400, thanks to a claimed energy
    resolution of about 1.5~\% at 100 GeV, will be able to separate a
    $\gamma\gamma$ line from a $Z\gamma$ or $H\gamma$, if the corresponding
    branching ratio is comparable to that into two photons, while HESS-II and
    CTA cannot separate them.  
  \item We investigate for which signal fluxes a signal arising from the
    annihilation into $\gamma\gamma$ can be distinguished from IB photons, and
    found that GAMMA-400 would be able to distinguish at a $95\%$CL between a
    gamma-ray line and IB photons if the 130 GeV feature is real, and we have
    identified the broader region of the parameter space where a
    discrimination is possible.
\end{itemize}

HESS-II will soon be operational and given the good performances foreseen for
the instrument in hybrid-mode, we stress that it should offer a quick
confirmation of the genuineness of the signal reported in \cite{weniger} (our
estimates are based on an exposure time of 50 hours, assuming intermediate
zenith angles), since this could provide on a short timescale an independent
observation with completely different background and systematic errors. As for
CTA, the actual construction of the array should start in 2015, and the first
data should realistically be available by 2018. 

In the case of GAMMA-400, the claimed improvement in energy and angular
resolution over Fermi-LAT make it an invaluable tool for dark matter searches.
We have demonstrated that it has an enormous potential in the detection and
discrimination of lines, despite the smaller effective area compares to the
NASA satellite, and we therefore strongly encourage this experimental effort.

\smallskip
\acknowledgments

We are grateful for discussions with Yvonne Bechereni, Torsten Bringmann,
James Buckley, Michael Kuhlen, Emmanuel Moulin, Michael Punch and
Hannes-S.~Zechlin. The research of LB~was carried out under Swedish Research
Council (VR) contract no.~621-2009-3915. GB acknowledges the support of the
European Research Council through the ERC Starting Grant {\it WIMPs Kairos}.
CW acknowledges partial support from the European 1231 Union FP7 ITN
INVISIBLES (Marie Curie Actions, PITN-GA-2011-289442). JC is research fellow
of the Royal Swedish Academy of Sciences financed by a grant of the K\&A
Wallenberg foundation.


\end{document}